\documentclass[preprint,authoryear,11pt]{elsarticle}
\usepackage{amsmath,amssymb,amsthm}
\usepackage{graphicx}
\usepackage{natbib}
\journal{\null}

\begin{document}
\begin{frontmatter}

\numberwithin{equation}{section}
\theoremstyle{plain}
\newtheorem{theorem}{Theorem}
\newtheorem{lemma}{Lemma}
\newtheorem{corollary}{Corollary}
\newtheorem{proposition}{Proposition}

\theoremstyle{definition}
\newtheorem*{definition}{Definition}
\newtheorem{example}{Example}
\newtheorem*{continue1}{Example 1 continued}
\newtheorem*{continue2}{Example 2 continued}

\theoremstyle{remark}
\newtheorem*{remark}{Remarks}

\title{Nonparametric Estimation of Trend in Directional Data}

\author{Rudolf Beran\fnref{ack}}

\address{Department of Statistics, University of California,
Davis,\\ One Shields Avenue, Davis CA 95616-8705, USA}

\fntext[ack]{This research was supported in part by National Science
Foundation Grant DMS-1127914 to the Statistics and Applied Mathematical
Sciences Institute, through the SAMSI program on Low-Dimensional Structure
in High-Dimensional Systems.}

\ead{rjberan@ucdavis.edu}

\begin{abstract} Consider measured positions of the paleomagnetic north
pole over time. Each measured position may be viewed as a direction,
expressed as a unit vector in three dimensions and incorporating some
error. In this sequence, the true directions are expected to be close to
one another at nearby times. A simple trend estimator that respects the
geometry of the sphere is to compute a running average over the
time-ordered observed direction vectors, then normalize these average
vectors to unit length. This paper treats a considerably richer class of
competing directional trend estimators that respect spherical geometry. The
analysis relies on a nonparametric error model for directional data in
$R^q$ that imposes no symmetry or other shape restrictions on the error
distributions. Good trend estimators are selected by comparing estimated
risks of competing estimators under the error model. Uniform laws of large
numbers, from empirical process theory, establish when these estimated
risks are trustworthy surrogates for the corresponding unknown risks.

\end{abstract}


\end{frontmatter}

\def\be{\begin{equation}}
\def\ee{\end{equation}}
\def\Mhat{\hat{M}}
\def\Dhat{\hat{D}}
\def\Rhat{\hat{R}}
\def\Rbreve{\breve{R}}
\def\Acal{{\cal A}}
\def\Ahat{\hat{A}}
\def\Atilde{\tilde{A}}
\def\Bbar{\bar{B}}
\def\atilde{\tilde{a}}
\def\ahat{\hat{a}}
\def\gammahat{\hat{\gamma}}
\def\mhat{\hat{m}}
\def\tauhat{\hat{\tau}}
\def\tautilde{\tilde{\tau}}
\def\what{\hat{w}}
\def\wtilde{\tilde{w}}

\def\E{\mathrm{E}}
\def\P{\mathrm{P}}
\def\Var{\mathop\mathrm{Var}}
\def\Cov{\mathop\mathrm{Cov}}
\def\Dif{\mathop\mathrm{Dif}}
\def\atan2{\mathop\mathrm{atan2}}
\def\argmin{\mathop\mathrm{argmin}}
\def\rank{\mathop\mathrm{rank}}
\def\tr{\mathop\mathrm{tr}}
\def\diag{\mathop\mathrm{diag}}
\def\plim{\mathop\mathrm{plim}}

\section{Introduction} \label{sec1} 
\subsection{Preliminaries} \label{subsec1.1}
Consider measurements on the position of the Earth's north magnetic pole,
derived from rock samples collected at various sites. Each observed
position, usually reported as latitude and longitude, may be represented as
a unit vector in $R^3$ that specifies the direction from the center of the
Earth to the point on the Earth's surface with that latitude and longitude.
Associated with each such direction vector is the geological dating of the
corresponding rock sample. Substantial measurement errors are to be
expected in the data. The problem is to extract trend in the position of
the north magnetic pole as function of time.    

Consider in $R^3$ the orthonormal basis $(j_1, j_2, j_3)$ in which $j_3$ is
the unit vector pointing to the Earth's geographical north pole and $j_1$
is the unit vector orthogonal to $j_3$ that points to longitude $0$. 
Relative to this basis, an observed direction has polar coordinates
$(\theta, \phi)$. Here $\theta \in [0, \pi]$ is the angle, in radians,
between $j_3$ and the observed direction. The angle $\phi \in [0, 2\pi)$
specifies, in radians, the counterclockwise rotation angle in the
$j_1$-$j_2$ plane from $j_1$ to the longitude of the observed direction. 

Relative to the same basis, the unit vector with polar coordinates
$(\theta, \phi)$ has Cartesian coordinates
\be
x_1 = \sin(\theta) \cos(\phi), \qquad x_2 = \sin(\theta) \sin(\phi), \qquad
x_3 = \cos(\theta).
\label{1.1}
\ee
Cartesian coordinates prove useful in defining trend estimators that
operate on directional data. From the Cartesian coordinates of a
direction, whether observed or fitted, the polar coordinates may be
recovered as 
\be
(\theta, \phi) =
\begin{cases}
 (\arccos(x_3),\atan2(x_2, x_1)) & \text{if $\atan2(x_2,x_1)\ge 0$} \\
(\arccos(x_3),\atan2(x_2, x_1) + 2\pi) & \text{otherwise}.
\end{cases}
\label{1.2}
\ee
The function $\atan2$, which has domain $R^2 - \{0,0\}$ and range $(-\pi,
\pi]$ is defined by
\be
\atan2(v,u) =
\begin{cases}
\arctan(v/u) & \text{if $u > 0$}\\
\arctan(v/u) + \pi & \text{if $u < 0, v \ge 0$}\\
\arctan(v/u) - \pi & \text{if $u < 0, v < 0$}\\
\pi/2 & \text{if $u = 0, v > 0$}\\
-\pi/2 & \text{if $u = 0, v < 0$}.
\end{cases}
\label{1.3}
\ee
Mathematical programming languages generally provide this function.

Important for visualizing directions in $R^3$ is the Lambert azimuthal
projection of a hemisphere. If a direction has polar coordinates $(\theta,
\phi)$, let 
\be (\rho, \psi) = \begin{cases} 
(2\sin(\theta/2),\phi) & \text{if $\theta \le \pi/2$}\\ 
(2\sin((\pi - \theta)/2,\phi) & \text{if $\theta > \pi/2$}. 
\end{cases} 
\label{1.4} 
\ee 
Then plot the direction as the projected point $(\rho\cos(\psi),
\rho(\sin(\psi))$ in $R^2$, using different plotting symbols according to
whether $\theta \le \pi/2$ (the northern hemisphere) or $\theta > \pi/2$
(the southern hemisphere). The first case, where $\theta \le \pi/2$, gives
an area preserving projection of the northern hemisphere into a disk of
radius $\sqrt 2$. The center of the disk represents the north pole and the
perimeter of the disk corresponds to the equator. The second case does
likewise for the southern hemisphere, the center of the disk now
representing the south pole. See Watson (1983) for a brief derivation of
the Lambert (or equal area) projection and a discussion of its use in
directional statistics.

\begin{figure}[ht]
\begin{center}
\includegraphics[height=2in]{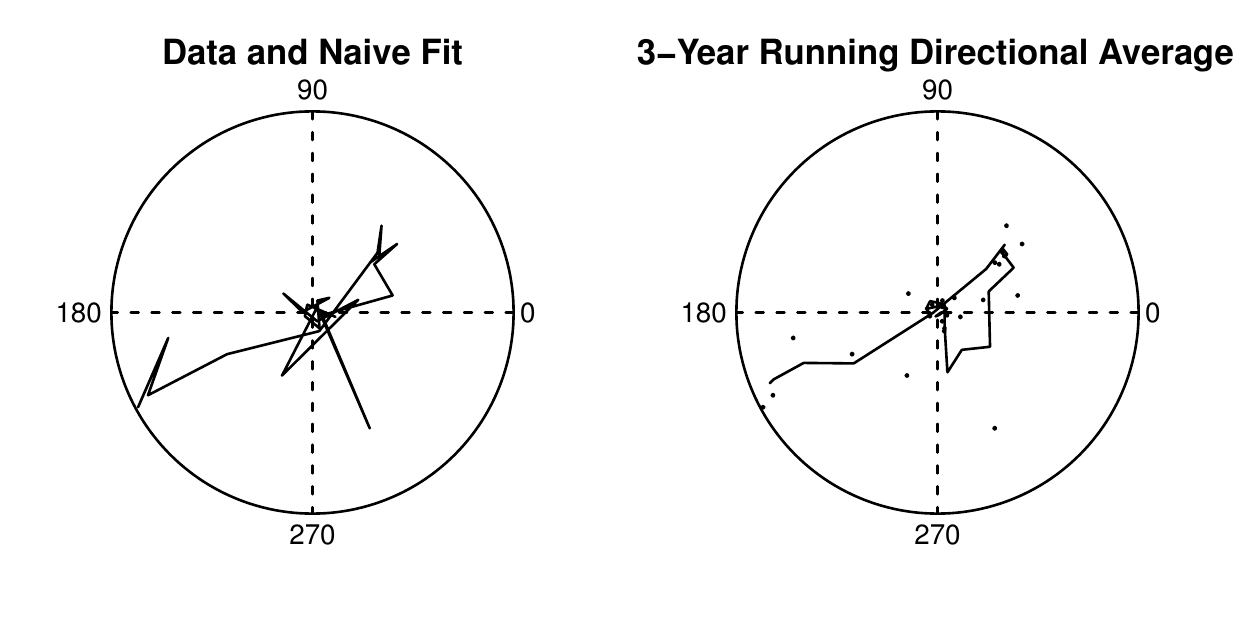}
\caption{\sl The measured positions of the paleomagnetic north magnetic
pole and two fits: interpolated raw data and interpolated 3-year moving
average direction. The interpolating lines merely indicate the time
sequence.}  
\label{Fig1}
\end{center}
\end{figure}

\citet[pp.\ 42--45]{J&K} reported 25 positions of the paleomagnetic north
pole, measured in rock specimens from various sites in Antarctica. Each
rock specimen was dated, so that the time sequence of the measured
positions is known. The left-hand Lambert plot in Fig.\ \ref{Fig1} displays
the measured magnetic pole positions, expressed in polar coordinates and
plotted according to \eqref{1.4}. Line segments join positions adjacent in
time. Little pattern emerges. Insight is gained by: (a) using \eqref{1.1}
to express each observed direction as a unit vector in Cartesian
coordinates; (b) forming a 3-year running average of these direction
vectors; (c) rescaling each average vector to unit length. Reflection
handles the end points in the sequence. Example \ref{eg1} in subsection
\ref{subsec1.2} gives details. After transformation \eqref{1.2} back to
polar coordinates, the right-hand Lambert plot in Fig.\ \ref{Fig1} joins
three-year running average directions adjacent in time with line
segments. Evident now is an apparent convergence, as time goes on, of the
paleomagnetic north pole to the geographic north pole. 

In their analysis of the data, \citet[Figs.\ 1 and 2]{J&K} unwrapped the
sphere and directional data onto a plane, used a continuous cubic spline
fit to the planar data, then wrapped this fit back onto the sphere. Their
continuous spline fit resembles the linearly interpolated 3-year moving
directional average except for a surprising kink near the left edge of the
Lambert plot. They remarked that this kink lacks physical significance, not
being supported by data, and is an artifact of their methodology. In
contrast, the running 3-year directional average avoids distortions caused
by projection of the sphere onto a plane.

Does the running 3-year average direction provide an effective estimate of
the trend in the paleomagnetic data? Clearly we cannot know from
a single data-analysis. In this paper, we assess the performance of
the three-year running directional average and of certain competing
estimators on hypothetical data described by workable probability models.
Because this approach examines performance of competing estimators over a
broad range of hypothetical data sets, it can reveal their strengths and
weaknesses. However, the relationship between data and probability
models for data is not untroubled. Insightful comments on the situation
include:
\begin{itemize}
\item 
\citet{JD}: ``The present formal correctness of mathematical
probability only helps indirectly in analyzing real probabilistic
phenomena. It is unnecessary to stress to statisticians that the relation
between mathematics and these phenomena is still obscure.''

\item
\citet{AK}: ``There arises a problem of finding the reasons for the
applicability of the mathematical theory of probability to the phenomena of
the real world.''

\item
\citet{JT}: ``In practice, methodologies have no assumptions and deliver
no certainties.'' 
\end{itemize}

\subsection{Outline of the Paper}\label{subsec1.2}
The model problem is to estimate an underlying trend from an observed
sequence $(y_1, y_2, \ldots y_p)$ of random column vectors that take values
in the unit sphere $S_q = \{y \in R^q \colon |y|=1\}$. Each of these
direction vectors is indexed by an ordinal covariate and is measured with
random error. Mean directions at nearby covariate values may be expected to
be close to one another. This paper develops a large class of estimators
for the mean directions and supporting theory. Key steps in the development
are as follows.

\begin{itemize}
\item{\sl Rescaled Symmetric Linear Estimators}. Let $Y$ be the $p \times
q$ data matrix with succesive rows $y_1', y_2', \ldots y_q'$. Let $A$ be a
symmetric $p \times p$ matrix. For any matrix $B$, let $\Delta(B) =
\diag(BB')$, the diagonal matrix obtained by replacing every off-diagonal
entry in $BB'$ with zero. Note that the matrix $\Delta^{-1/2}(B)B$
rescales each row of $B$ to unit length. Consider the estimators
\be
\Mhat(A) = AY, \qquad \Dhat(A) = \Delta^{-1/2}(AY)AY.
\label{1.5}
\ee
Under the probability model for $Y$ to be described later, $\Mhat(A)$ is
an estimator, usually biased, for the mean matrix $M = \E(Y)$. The rows of
$\Dhat(A)$, having unit length, are the implied estimators of the
directional means. 

\begin{example}\label{eg1}
Let
\be
A = \begin{pmatrix}
2/3&1/3&0&0& \ldots &0&0&0\\
1/3&1/3&1/3&0& \ldots &0&0&0\\
0&1/3&1/3&1/3& \ldots &0&0&0\\
\vdots&\vdots&\vdots&\vdots& \ddots &\vdots&\vdots&\vdots\\
0&0&0&0& \ldots &1/3&1/3&1/3\\
0&0&0&0& \ldots &0&1/3&2/3
\end{pmatrix}
\label{1.6}
\ee
Then $\Dhat(A)$ is the 3-year running directional average described in the
previous subsection. Evidently, the construction generalizes to weighted
running averages of any odd-numbered span. Section \ref{sec3} treats these
and further classes of matrices $A$ useful for estimation of directional
trend.
\end{example}

\item{\sl Selection Procedure for $A$}.
Section \ref{sec2} presents a general nonparametric probability model for
the matrix $Y$ of directional measurements. The performance of directional
mean estimator $\Dhat(A)$ will be assessed through its risk computed under
this model. Let $|\cdot|$ denote the Frobenius matrix norm:
$|B|^2 = \tr(BB') = \tr(B'B)$. The \textsl{extrinsic loss} of $\Dhat(A)$ is
defined to be 
\be
L(\Dhat(A), M) = p^{-1}|\Mhat(A) - M|^2.
\label{1.7}
\ee
This choice of loss is algebraically tractable for subsequent theory and
will be seen to work reasonably on artificial and real directional data.

Because the corresponding \textsl{extrinsic risk} of $\Dhat(A)$ depends on
unknown parameters, it is necessary to estimate it from $Y$.  As will be
seen in section \ref{sec2}, the estimated risk can be expressed as
\be
\Rhat(A) = p^{-1}[|Y - AY|^2 + (2\tr(A) - p)\gammahat^2],
\label{1.8}
\ee
where $\gammahat^2$ is an estimator of a dispersion parameter $\gamma^2$ to
be defined in \eqref{2.5}. Equation \eqref{1.8}, for which an asymptotic
justification will be provided, is related to some considerations in
\citet{CM}.

Candidate estimators $\{\Dhat(A) \colon A \in \Acal\}$ are compared through
their estimated risks. Of particular potential is the candidate estimator
with smallest estimated risk:
\be
\Dhat(\Ahat), \quad \mathrm{\ where\ }\Ahat = \argmin_{A \in
\Acal}\Rhat(A).
\label{1.9}
\ee

\item{\sl Role of Uniform Laws of Large Numbers}.
Suppose that $\Acal$ is a rich class of $p \times p$ symmetric matrices
constructed to respect prior vague hypotheses about the mean directions.
Suppose that $\Atilde$ minimizes the risk of $\Dhat(A)$ over all $A \in
\Acal$. When does the risk of the adaptive directional estimator
$\Dhat(\Ahat)$ defined above approximate closely the risk of the
ideal unrealizable estimator $\Dhat(\Atilde)$? 

An asymptotic answer, in which the number $p$ of directions observed tends
to infinity, draws on uniform laws of large numbers. Let $W(A)$ denote
either the loss or the estimated risk $\Rhat(A)$ of directional estimator
$\Dhat(A)$. Suppose that, as $p$ increases, $W(A)$ converges in
$L_1$ norm to the risk of $\Dhat(A)$, {\sl uniformly over all} $A \in
\Acal$. Then the risk of the adaptive estimator $\Dhat(\Ahat)$ converges to
the risk of ideal estimator $\Dhat(\Atilde)$. Moreover, the plug-in risk
estimator $\Rhat(\Ahat)$ also converges to the risk of $\Dhat(\Ahat)$.
These convergences support use of estimated risk to score the adequacy of
the candidate estimator class considered. Section \ref{sec3} presents
theorems of this type, in which empirical process theory plays a central
role. Section \ref{sec5} gives all theorem proofs.

\item{\sl Numerical Experiments}.
Examples in Section \ref{sec4} illustrate how vague prior hypotheses about
mean directions combined with estimator selection through estimated risks
can generate effective estimators of mean directions. The experiments treat
three artificial data sets where the truth is known, then analyze further
the paleomagnetic north pole data.

\end{itemize}

\subsection{Perspectives}\label{subsec1.3} 
Other directional trend estimators proposed by \citet{GW2}, \citet{F&L},
and \citet{J&K} are curve fitting procedures that rely on analogs of
cubic-spline or kernel methods in Euclidean spaces. Important in such
treatments are assumptions on the continuity and smoothness of the unknown
trend. The estimators of this paper neither fit nor assume an underlying
smooth trend. They are discrete estimators of discrete mean directions that
take advantage of any underlying slow variation to reduce estimation risk
and perform accountably in any case. 

Running average directions are a special case of a more general concept,
running extrinsic Fr\'echet averages. For definitions and some applications
of Fr\'echet means and averages, see \citet{B&P}. The developments in this
paper are potentially extensible to estimating trends in Fr\'echet means
through running Fr\'echet averages and, more generally, through analogs of
\eqref{1.5}.

For instance, representing axes as projection matrices of rank one enables
construction of running average axes. An average axis is defined to be the
rank-one projection matrix onto the eigenvector associated with the largest
eigenvalue of the average of the observed projection matrices. This
definition is simply the Euclidean projection of the averaged rank-one
projection matrices into the space of rank-one projection matrices. See
\citet{B&F} for details. A richer class of candidate estimators for trend
in axes is obtained by modifying \eqref{1.5} so that each
row of $Y$ is now a vectorized observed rank-one projection. Each row of
$\Mhat(A) = AY$, reassembled into a symmetric matrix, is then projected, in
Euclidean metric, into the set of projection matrices of rank one. The
result defines $\Dhat(A)$ for axes.

\section{Nonparametric Model, Risks, and Estimated Risks}\label{sec2}
This section introduces a nonparametric model for the matrix $Y$ of
observed directions in $R^q$. The quadratic risk of candidate estimator
$\Dhat(A)$ is calculated under this model. Estimated risk is then derived.

\subsection{Data Model}\label{subsec2.1}
Let $\{e_i \colon 1 \le i \le p\}$ be independent identically distributed
random unit vectors in $S_q$ such that
\be
\E(e_i) = \lambda\mu_0, \mathrm{\ where\ } \mu_0 = (1,0,0,\ldots 0)'
\mathrm{\ and\ }\lambda \ne 0.
\label{2.1}
\ee
The distribution of $e_i$ and the value of $\lambda$ are otherwise unknown.
No shape assumptions are made on the former. The covariance matrix of $e_i$
then satisfies
\be
\Sigma = \Cov(e_i) = \E(e_ie_i') - \lambda^2 \mu_0\mu_0', \qquad \gamma^2 =
\tr(\Sigma) = 1 - \lambda^2 \ge 0.
\label{2.2}
\ee
Thus, $0 <|\lambda| \le 1$, in agreement with geometrical intuition.

The data model specifies we observe $y_i = O_ie_i$ for $1 \le i
\le p$, where each $O_i$ is an unknown nonrandom $q \times q$ orthogonal
matrix. The $\{y_i\}$ are independent random unit vectors in $S_q$ with
mean vectors and covariance matrices that depend on $i$:
\be
m_i = \E(y_i) = \lambda O_i\mu_0, \qquad \Cov(y_i) = O_i\Sigma O_i'.
\label{2.3}
\ee
The unit vector $\mu_i = O_i\mu_0$ is the {\sl mean direction} of $y_i$.
Because $\lambda \ne 0$, $\mu_i = m_i/||m_i|$.

The $p \times q$ {\sl data matrix} $Y = \{y_{ij}\}$ has rows $y_1', y_2',
\ldots y_p'$, each of unit length, and columns $y_{(1)}, y_{(2)}, \ldots
y_{(q)}$. The corresponding {\sl mean matrix} $M = \E(Y)$ has rows $m_1',
m_2', \ldots m_p'$ and columns $m_{(1)}, m_{(2)}, \ldots m_{(q)}$. In
particular, $m_{(j)} = \E y_{(j)}$. Because the elements of $y_{(j)}$ are
independent random variables,
\be
\Cov(y_{(j)}) = \diag\{\Var (y_{ij}) \colon 1 \le i \le p\}.
\label{2.4}
\ee
In view of \eqref{2.3},
\be
\sum_{j=1}^q \Var (y_{ij}) = \tr (\Cov(y_i)) = \tr(O_i \Sigma O_i') =
\tr(\Sigma) = \gamma^2, \qquad 1 \le i \le p.
\label{2.5}
\ee
Equations \eqref{2.4} and \eqref{2.5} imply
\be
\sum_{j=1}^q \Cov(y_{(j)}) = \gamma^2 I_p, \qquad 
\E (\sum_{j=1}^q y_{(j)}y_{(j)}') =\sum_{j=1}^q m_{(j)}m_{(j)}'+\gamma^2
I_p.
\label{2.6}
\ee
The calculations of risk and quadratic risk in the next subsection draw on
these expressions.

\subsection{Loss and Risk}\label{subsec2.2}
For the candidate estimator $\Dhat(A)$ defined in \eqref{1.5}, the
quadratic {\sl loss} and {\sl risk} are taken to be 
\be
L(\Dhat(A),M) = p^{-1}|AY - M|^2, \quad R(\Dhat(A),M,\gamma^2)= \E
L(\Dhat(A),M).
\label{2.7}
\ee
The next theorem shows how the risk depends on the model only through $M$
and $\gamma^2$.

\begin{theorem}\label{thm1}
Under the nonparametric data model of section \ref{sec2}, the risk of the
candidate estimator $\Dhat$ is
\be
R(\Dhat(A),M,\gamma^2) = p^{-1}[\gamma^2\tr(A^2) + \tr((I_p - A)^2MM')].
\label{2.8}
\ee
Suppose that the $p \times p$ symmetric matrix $A$ has spectral
representation 
\be
A = \sum_{k=1}^s a_kP_k, \qquad s \le p,
\label{2.9}
\ee
where the $\{a_k\}$ are the distinct eigenvalues and the $\{P_k\}$ are the
corresponding mutually orthogonal eigenprojections. Then
\begin{align}
R(\Dhat(A),M,\gamma^2) &= \sum_{k=1}^s [a_k^2 \tau_k + (1 - a_k)^2
w_k]\notag\\
&= \sum_{k=1}^s [(a_k - \atilde_k)^2(\tau_k + w_k) + \tau_k\atilde_k],
\label{2.10}
\end{align}
where $\tau_k = p^{-1}\gamma^2\tr(P_k)$, $w_k = p^{-1}|P_kM|^2$, and
$\atilde_k = w_k/(\tau_k + w_k)$. For fixed $\{P_k\colon 1 \le k \le s\}$,
the risk $R(\Dhat(A),M,\gamma^2)$ is minimized when $a_k = \atilde_k$ for
every $k$.
\end{theorem}
Because $\atilde \in [0,1]$, it follows that only symmetric matrices $A$
whose eigenvalues all lie in $[0,1]$ need to be considered when seeking to
minimize the risk of $\Dhat(A)$. Such matrices accomplish Stein-type oracle
shrinkage in each of the eigenspaces.

\subsection{Estimated Risk}\label{subsec2.3}
Estimating the risk $R(\Dhat(A),M,\gamma^2)$ in \eqref{2.8} requires
estimating $\gamma^2$ and $MM' = \sum _{j=1}^q m_{(j)}m_{(j)}'$. It follows
from \eqref{2.6} that $\E(YY') = MM'+ \gamma^2 I_p$. Consequently the naive
estimator $p^{-1}\tr((I_p - A)^2YY')$ for the risk term $p^{-1}\tr((I_p -
A)^2MM')$ in \eqref{2.8} is biased upward by the amount $p^{-1}\tr((I_p -
A)^2\gamma^2)$. In general, the bias does not vanish as $p$ increases. 

For $\gamma^2$, consider the estimator 
\be 
\gammahat^2 = [2(p-1)]^{-1} \sum_{i=2}^2 |y_i - y_{i-1}|^2. 
\label{2.11} 
\ee 
It will seen in section \ref{sec3} that, as $p$ increases, $\gammahat^2$
converges in $L_1$ norm to $\gamma^2$ provided $p^{-1}\sum _{i=2}^2 |m_i -
m_{i-1}|^2$ tends to zero. This makes $\gammahat^2$ a reasonable estimator
of $\gamma^2$ when successive mean directions vary slowly. Other estimators
of $\gamma^2$ may be constructed by using higher order differences of the
$\{y_i\}$ or taking advantage of replication in the data set. Given a
consistent estimator $\gammahat^2$ of $\gamma^2$, the obvious bias
correction yields the {\sl estimated risk} 
\be 
\Rhat(A) = p^{-1}[\gammahat^2 \tr(A^2) + \tr((I_p - A)^2(YY' -
\gammahat^2I_p))] 
\label{2.12} 
\ee

\begin{theorem}\label{thm2}
The estimated risk \eqref{2.12} of $\Dhat(A)$ has the alternative
expression
\be
\Rhat(A) = p^{-1}[|Y-AY|^2 + (2\tr(A) - p)\gammahat^2].
\label{2.13}
\ee
Suppose the symmetric matrix $A$ has the spectral representation
\eqref{2.9}. Then
\begin{align}
&\Rhat(A) = \sum_{k=1}^s [a_k^2 \tauhat_k + (1 - a_k)^2 \what_k]\notag\\
&= \sum_{\what_k \ge 0} [(a_k - \ahat_k)^2(\tauhat_k + \what_k) +
\tauhat_k\ahat_k] + \sum_{\what_k < 0} [a_k^2 \tauhat_k + (1 - a_k)^2
\what_k] ,
\label{2.14}
\end{align}
where $\tauhat_k = p^{-1}\gammahat^2\tr(P_k)$, $\what_k = p^{-1}|P_kY|^2 -
\tauhat_k^2$, and $\ahat_k = \what_k/(\tauhat_k + \what_k)$ if $\what_k \ge
0$ but $\ahat_k = 0$ if $\what_k < 0$. For fixed $\{P_k \colon 1 \le k \le
s\}$, the estimated risk $\Rhat(A)$ is minimized when $a_k = \ahat_k$ for
every $k$.
\end{theorem}
The differences between the definitions of $\ahat_k$ here and of
$\atilde_k$ in the preceding theorem arise because, unlike $w_k$,
the estimated quantity $\what_k$ can take on negative values.

For a fixed set of $\{P_k \colon 1 \le k \le s\}$, the symmetric matrix
$\Ahat = \sum_{k=1}^s \ahat_kP_k$ minimizes estimated risk over the
candidate class of matrices $A$ defined by \eqref{2.9}. The oracle
symmetric matrix $\Atilde = \sum_{k=1}^s \atilde_kP_k$ minimizes risk over
the same candidate class of $A$.  The trustworthiness of $\Ahat$ as a
surrogate for the unrealizable $\Atilde$ requires investigation. Key is
that the estimated risk $\Rhat(A)$ converge to the risk
$R(\Dhat(A),M,\gamma^2)$ uniformly over the class $\Acal$ of symmetric
matrices $A$ under consideration as $p$ tends to infinity. This is a
question in empirical process theory, addressed in Section \ref{sec3}. For
an example of a simpler situation where uniform convergence fails and
$\Ahat$ specifies a poor estimator, see Remark A on p.\ 1829 of
\citet{B&D}.

The estimated risk $\Rhat(A)$ can be negative. This does not affect the
rationale for ranking competing estimators according to the order of their
estimated risks. Adding a very large constant to the loss function reduces
the chances that estimated risk is negative without changing anything
essential. 

\section{Uniform Convergence of Estimated Risks and Adaptation}\label{sec3}
Uniform laws of large numbers for estimated risks are the main theme of
this section. These yield an asymptotic justification for adaptive
estimation of mean direction by minimizing estimated risk. Consistent
estimation of the dispersion parameter $\gamma^2$ is treated. 

Many quantities arising in the discussion depend on $p$. For readability,
this dependence is usually suppressed in the notation.

\subsection{Assumptions}\label{subsec3.1}
For any matrix $B$, let $|B|_{sp} = \sup_{x \ne 0}[|Bx|/|x|]$ denote
its spectral norm. The candidate estimators of mean direction are
$\{\Dhat(A) \colon A \in \Acal\}$, as defined in \eqref{1.5}. The following
assumptions are made:
\begin{itemize}
\item
$\Acal = \{A(t) \colon t \in [0,1]^k, A(t)\mathrm{\ is\ } p\times p 
\mathrm{\ symmetric}\}$.

\item

$A(t)$ is continuous on $[0,1]^k$ with $\sup_p\sup_{t \in [0,1]^k}
|A(t)|_{sp} < \infty$.

\item
$A(t)$ is differentiable on the interior of $[0,1]^k$, with partial
derivatives $\{\nabla_i A(t) = \partial A(t)/\partial t_i \colon 1 \le i
\le k\}$. The partial derivatives satisfy $\sup_{p,i}\sup_{t \in [0,1]^k}
|\nabla_i A(t)|_{sp} < \infty$.

\item
The data matrix $Y$ satisfies the probability model described in subsection
\ref{subsec2.1}.

\item
The estimator $\gammahat^2$ of $\gamma^2$ is uniformly $L_1$ consistent:
\be
\lim_{p \rightarrow \infty} \sup_M \E |\gammahat^2 - \gamma^2| = 0.
\label{3.1}
\ee

\end{itemize}
Candidate estimators that meet the assumptions on $\Acal$ arise naturally.
Two examples illustrate:

\begin{continue1}{\sl Running weighted average directions} The span-3
running average directions of \eqref{1.6} can be generalized to span-3
running weighted average directions, defined by the class of candidate
estimators generated by
\be
A(t) = \begin{pmatrix}
t_1 + t_2&t_2&0&0& \ldots &0&0&0\\
t_2&t_1&t_2&0& \ldots &0&0&0\\
0&t_2&t_1&t_2& \ldots &0&0&0\\
\vdots&\vdots&\vdots&\vdots& \ddots &\vdots&\vdots&\vdots\\
0&0&0&0& \ldots &t_2&t_1&t_2\\
0&0&0&0& \ldots &0&t_2&t_1 + t_2
\end{pmatrix}
\label{3.2}
\ee
for $t = (t_1, t_2) \in [0,1]^2$ such that $t_1 + 2t_2 = 1$. The first and
last rows are obtained by reflection at the boundary. Because $|A(t)x|^2
\le [(t_1 + t_2)^2 + t_2^2]|x|^2$, it follows that $|A(t)|_{sp} \le
5^{1/2}$ for every $t \in
[0,1]^2$. Symmetrically weighted running average directions were already
used by \citet{EI} in a data analysis of continental drift data. Note that
$\nabla_1 A(t) = I_p$ and
\be
\nabla_2 A(t) = \begin{pmatrix}
1&1&0&0& \ldots &0&0&0\\
1&0&1&0& \ldots &0&0&0\\
0&1&0&1& \ldots &0&0&0\\
\vdots&\vdots&\vdots&\vdots& \ddots &\vdots&\vdots&\vdots\\
0&0&0&0& \ldots &1&0&1\\
0&0&0&0& \ldots &0&1&1
\end{pmatrix}.
\label{3.3}
\ee
Thus, $\sup_{t \in [0,1]^2}|\nabla_1 A(t)|_{sp} = 1$ and $\sup_{t \in
[0,1]^2}|\nabla_2 A(t)|_{sp} \le 2^{1/2}$. 
\end{continue1}

\begin{example}\label{eg2}{\sl Multiply penalized least squares}. Penalized
least squares estimators for Euclidean means inspire the following
development for directional means. Let $t = \{t_i \colon 1 \le i
\le k\}$ be penalty weights such that $0 \le t_i \le 1$. Let $c$ be a
positive constant, possibly very large. Let $\{Q_i\colon 1 \le i \le k\}$
be symmetric, positive semi-definite matrices, normalized so that
$|Q_i|_{sp} = 1$ for $1 \le i \le k$. Consider the class of candidate
directional mean $\Dhat(A(t))$ estimators generated by 
\be
A(t) = (I_p + Q(t))^{-1} , \quad{\rm where\ } Q(t) = c\sum_{i=1}^k t_iQ_i,
\quad t \in [0,1]^k.
\label{3.4}
\ee
Note that $A(t)Y$ is the value of $M$ that minimizes the penalized least
squares (PLS) criterion $|Y-M|^2 + \tr(M'Q(t)M)$.

Let $B(t)= A^{-1}(t)$. The eigenvalues of $B(t)$ all lie in $[1,1 + ck]$
for every $t \in [0,1]^k$. Consequently, $\sup_{t \in [0,1]^k}|A(t)|_{sp}
\le 1$. Moreover
\be
\nabla_i A(t)= -B^{-1}(t)[\nabla_i B(t)]B^{-1}(t) =
cB^{-1}(t)Q_i B^{-1}(t).
\label{3.5}
\ee
Thus,
\be
|\nabla_i A(t)|_{sp} \le c |B^{-1}(t)|_{sp}|Q_i|_{sp} |B^{-1}(t)|_{sp}
\le c(1 + ck)^{-2} \quad \forall t \in [0,1]^k.
\label{3.6}
\ee
\end{example}

\subsection{Uniform Laws of Large Numbers and Adaptation}\label{subsec3.2}
The uniform laws of large numbers in this section hold for matrices in the
class $\Acal = \{A(t) \colon t \in [0,1]^k\}$ satisfying the assumptions of
the preceding subsection. 

\begin{theorem}\label{thm3} Suppose that the assumptions of subsection
\ref{subsec3.1} hold. Let $W(A)$ denote either the loss $L(\Dhat(A),M)$ or
the estimated risk $\Rhat(A)$ of candidate directional estimator
$\Dhat(A)$.
Then, for every finite $\gamma^2 > 0$,
\be
\lim_{p \rightarrow \infty} \sup_M \E[\sup_{A \in \Acal} |W(A) -
R(\Dhat(A),M,\gamma^2)|] = 0.
\label{3.7}
\ee
\end{theorem}

Theorem \ref{thm3} asserts that the loss, risk, and estimated risk of
candidate estimator $\Dhat(A)$ converge together, uniformly over all $A \in
\Acal$, as the number $p$ of mean directions observed increases. 
As in the Introduction, let
\be
\Ahat = \argmin_{A \in \Acal}\Rhat(A), \qquad \Atilde = \argmin_{A \in
\Acal}R(\Dhat(A), M, \gamma^2) . 
\label{3.8}
\ee
The uniform convergence \ref{3.7} justifies using the {\sl adaptive}
directional estimator $\Dhat(\Ahat)$ as a surrogate for the unrealizable
oracle estimator $\Dhat(\Atilde)$, which achieves minimal risk over the
candidate class $\{\Dhat(A) \colon A \in \Acal\}$. 

\begin{theorem}\label{thm4} Suppose that the assumptions of subsection
\ref{subsec3.1} hold. Then, for every finite $\gamma^2 > 0$,
\be
\lim_{p \rightarrow \infty} \sup_M |R(\Dhat(\Ahat),M,\gamma^2) -
R(\Dhat(\Atilde),M,\gamma^2)| = 0.
\label{3.9}
\ee
Moreover, for $V$ equal to either the loss or risk of the adaptive
directional estimator $\Dhat(\Ahat)$,
\be
\lim_{p \rightarrow \infty} \sup_M \E|\Rhat(\Ahat) - V| =
0.
\label{3.10}
\ee
\end{theorem}

Theorem 4 shows that the risk of the adaptive directional estimator
$\Dhat(\Ahat)$ converges to the best possible risk achievable over the
class of candidate estimators $\{\Dhat(A)\colon A \in \Acal \}$. Equation
\eqref{3.10} establishes that the plug-in estimated risk $\Rhat(\Ahat)$
converges to the loss or risk of $\Dhat(\Ahat)$. This makes estimated risk
a credible metric for scoring the relative performance of competing
candidate
directional estimators. 

\subsection{Consistent Estimation of $\gamma^2$}\label{subsec3.3}
In the situation treated in this paper, where there one noisy observation
on each mean direction, estimating $\gamma^2$ consistently requires further
model additional assumptions. The next theorem shows how the presence of
slow variation between successive unknown mean directions can be used to
estimate $\gamma^2$. Analogous results hold for estimators of $\gamma^2$
based on higher order differences of the observed directions.

\begin{theorem}\label{thm5} 
Suppose that the assumptions of subsection \ref{subsec3.1} prior to
\eqref{3.1} hold and that
\be
\lim_{p \rightarrow\infty}[2(p-1)]^{-1}\sum_{i=2}^p |m_i - m_{i-1}|^2 - 0.
\label{3.11}
\ee
Let
\be
\gammahat^2 = [2(p-1)]^{-1}\sum_{i=2}^p |y_i - y_{i-1}|^2 .
\label{3.12}
\ee
Then, for every finite $\gamma^2 > 0$,
\be
\lim_{p \rightarrow \infty} \sup_M \E |\gammahat^2 - \gamma^2| = 0.
\label{3.13}
\ee

\end{theorem}

\section{Numerical Experiments}\label{sec4}
This section compares, on artificial directional data in three dimensions,
two natural adaptive penalized least squares (PLS) directional mean
estimators and the span-3 running directional average described in the
Introduction. The adaptive PLS estimators are then applied to the
paleomagnetic north pole data described in the Introduction.

\subsection{Candidate $k$-th difference PLS Estimators} \label{subsec4.1}
Consider the $(g-1) \times g$ first-difference matrix $\Dif(g) =
\{\delta_{u,v}\}$ in which $\delta_{u,u} = 1, \delta_{u,u+1} = -1$ for
every $u$ and all other elements are $0$. The $(p-1) \times p$
first-difference matrix is $\Delta_1 = \Dif(p)$. Define the $(p-d) \times
p$ \textsl{ $d$-th difference} matrix recursively through
\be
\Delta_1 = \Dif(p),\quad \Delta_d = \Dif(p-d+1)\Delta_{d-1},\quad 2 \le d
\le p-1
\label{4.1}
\ee

The $d$-th difference PLS directional mean estimator $\Dhat(A(t))$ is
defined by a special case of equation \eqref{3.4}, in which
\be
A(t) = (I_p + ct\Delta_d'\Delta_d/|\Delta_d'\Delta_d|_{sp})^{-1} , 
\quad t \in [0,1].
\label{4.2}
\ee
This PLS fit nudges the raw data $Y$ towards a smoother fit, according to
the choice of the difference-order $d$ in the penalty term and the
magnitude of the penalty weight $ct > 0$.

\subsection{Generating Artificial Directional Trend Data}\label{subsec4.2}
Let $U_1$, $U_1$ be independent random variables, each uniformly
distributed on $[0,1]$. Let
\be
\delta = \log[1 + (\exp(2\kappa) -1) U_1],\quad  \theta =
\cos^{-1}(\delta/\kappa -1),\quad  \phi = 2\pi U_2.
\label{4.3}
\ee
The random unit vector $z$ in $R^3$ with polar coordinates $(\theta, \phi)$
has a Fisher-Langevin distribution with mean direction $\nu_0 = (0,0,1)'$
and precision $\kappa$ (cf.\ \cite{M&J}). For any unit vector
$\mu$, the $3 \times 3$ orthogonal matrix
\be
\Omega(\mu) = (1 + \nu_0'\mu)^{-1}(\nu_0 + \mu)(\nu_0 + \mu)' - I_3
\label{4.4}
\ee
rotates $\nu_0$ into $\mu$ (cf.\ \citet[p.\ 28]{GW1}). Hence the random
vector $\Omega(\mu)z$ has a Fisher-Langevin distribution with mean
direction $\mu$ and precision $\kappa$.

Let $f\colon [0,1] \rightarrow [0,\pi]$ and $g \colon [0,1] \rightarrow
[0,2\pi)$ be designated functions. The pairs 
\be
\theta_i = f(i/(p+1)), \quad \phi_i = g(i/(p+1)), \quad 1 \le i \le p
\label{4.5}
\ee
define, in polar coordinates, a sequence of unit vectors $\{\mu_i \colon 1
\le i \le p\}$. Let $\{z_i \colon 1 \le i \le p\}$ be independent random
unit vectors, each constructed as in the preceding paragraph to have a
Fisher-Langevin distribution with mean direction $\nu_0$ and precision
$\kappa$. Then the random unit vectors
\be
y_i = \Omega(\mu_i)z_i
\label{4.6}
\ee
are independent and $y_i$ has a Fisher-Langevin distribution with mean
direction $\mu_i$ and precision $\kappa$. This construction is a
special case of the nonparametric data model defined in subsection 2.1. The
Fisher-Langevin model has a historical record of fitting some small
directional data sets plausibly (cf.\ \citet{F&L&E}, \citet{M&J}). 

\subsection{Numerical Trials on Artificial Data} \label{subsec4.3}
Figures 2, 3, and 4 exhibit competing directional trend fits to three
artificial data sets constructed as described in subsection \ref{4.2}.
These are compared with the true mean directions in each case. In
each data set, $p = 150$ directional means $\{\mu_i \}$ are observed with
error. The observed directions $\{y_i \colon 1 \le i \le 150\}$ are
pseudorandom unit vectors such that $y_i$ has a Fisher-Langevin
distribution with mean direction $\mu_i$ and precision $\kappa = 200$.
The functions $f$ and $g$ that define, through \eqref{4.5}, the polar
coordinates for each of the three sets of true directional means
$\{\mu_i\}$ are:

\begin{itemize}
\item
\textsl{Wobble:} $f(t) = .3\pi(t+.2 +.15\sin(36\pi t))$ while $g(t) = 4\pi
t$.

\item 
\textsl{Bat:} $f(t) = .8\pi (t-.5)$ while $g(t) = 4\pi \sin(6\pi t)$.

\item
\textsl{Jumps:} $f(t) = .2\pi$ if $0 \le t \le .15$, $= .1\pi$ if $.15 < t
\le .3$, $= .4\pi$ if $.3 < t \le .45$, $= .2\pi$ if $.45 < t \le .65$, $=
.3\pi$ if $.65 < t \le .8$, and $= .4 \pi$ if $.8 < t \le 1$ while $g(t) =
2\pi t$.

\end{itemize}

\begin{figure}[ht]
\begin{center}
\includegraphics[height=3in]{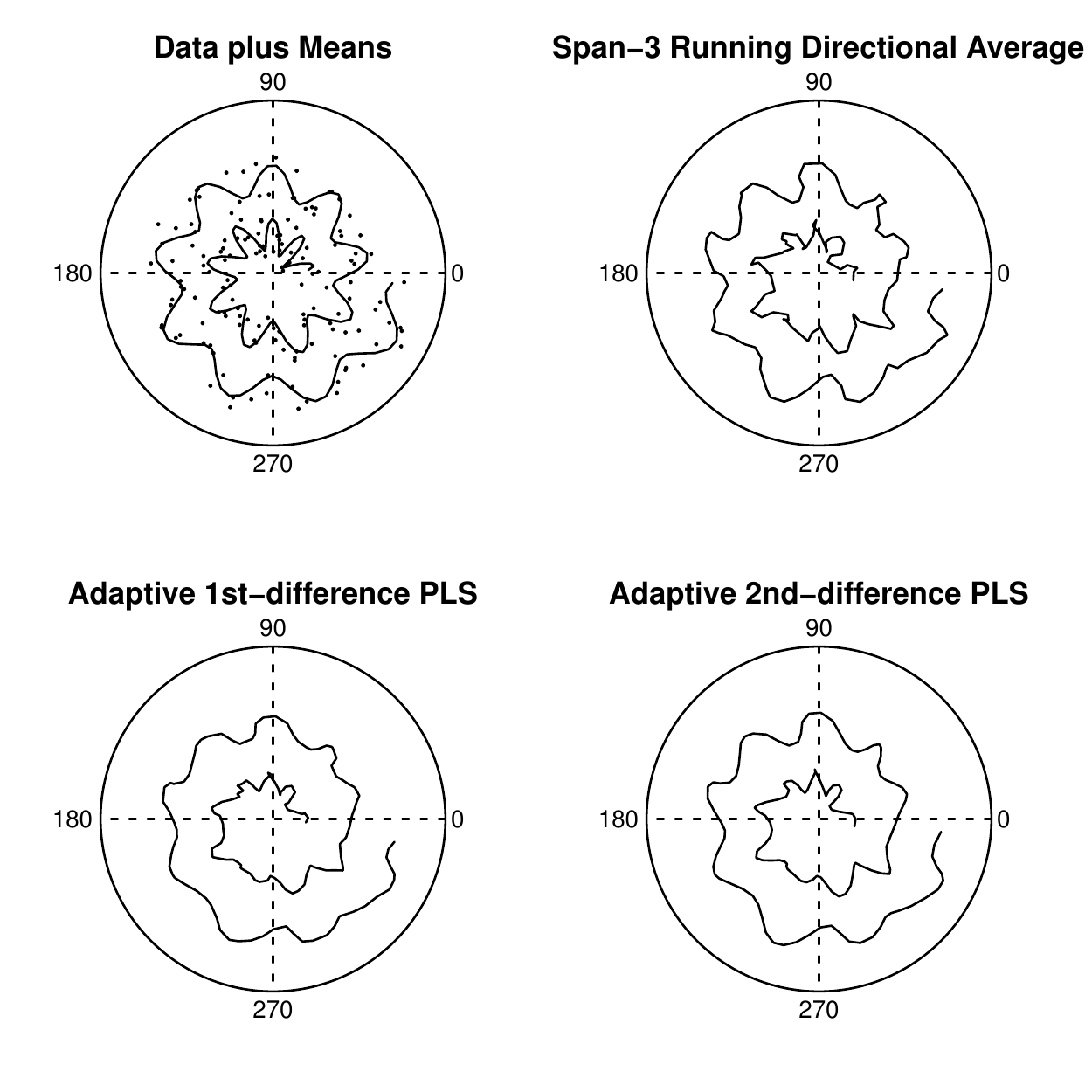}
\caption{\sl The Wobbly Spiral artificial data, together with the
underlying true mean directions, plus three competing fits to the data.
Interpolating lines indicate the time sequence of the true and the
estimated mean directions.}  
\label{Fig2}
\end{center}
\end{figure}

The artificial Wobble data are inspired by observations on the
Chandler-wobble of the geographic north pole, scaled up to wander over
a larger part of the northern hemisphere and given greater measurement
errors. \citet{DB} analysed actual Chandler-wobble data by using
time-series methods in the tangent plane of the sphere at the north pole.

\begin{figure}[ht]
\begin{center}
\includegraphics[height=3in]{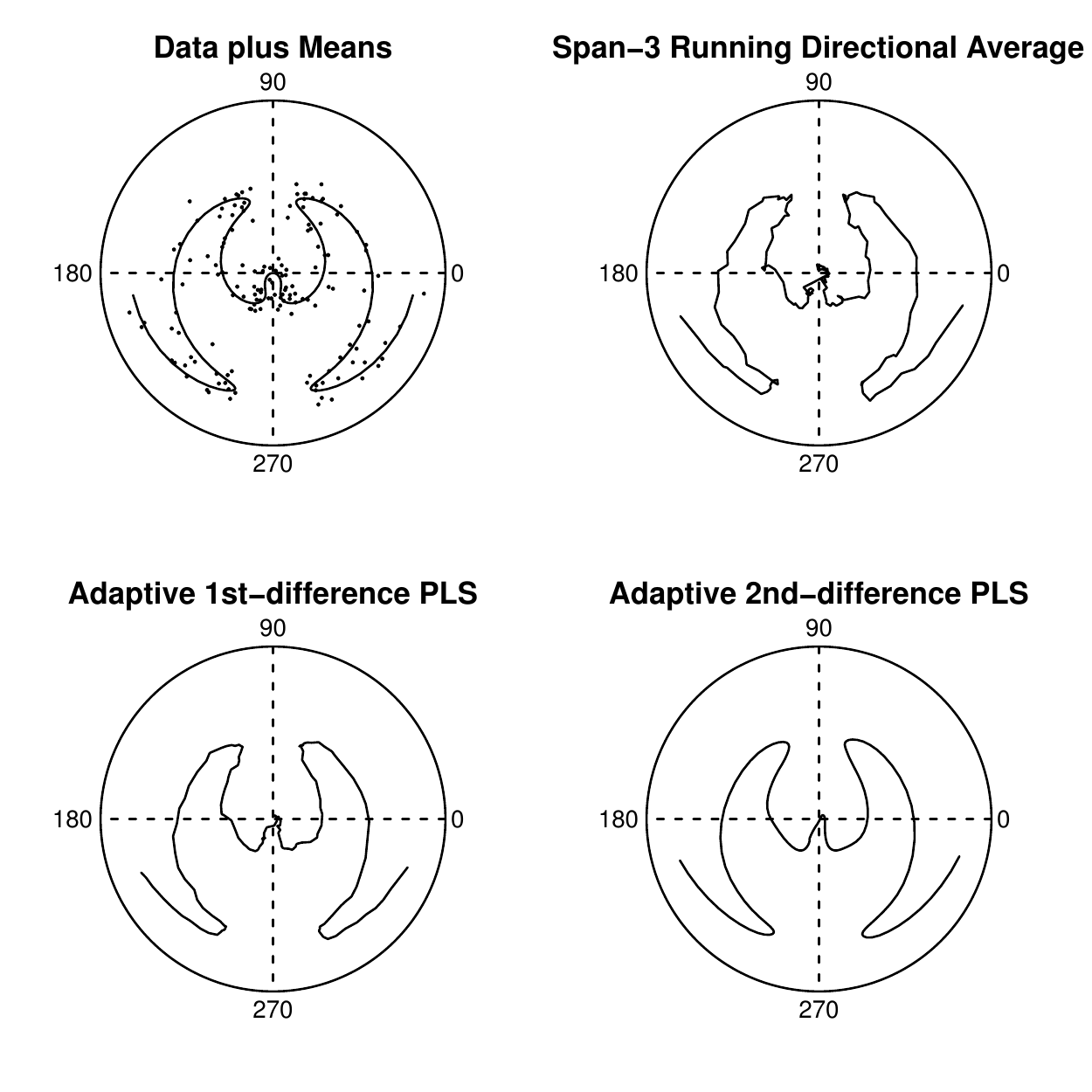}
\caption{\sl The Bat artificial data, together with the underlying true
mean directions, plus three competing fits to the data. Interpolating lines
indicate the time sequence of the true and estimated mean directions.}  
\label{Fig3}
\end{center}
\end{figure}

\begin{figure}[ht]
\begin{center}
\includegraphics[height=3in]{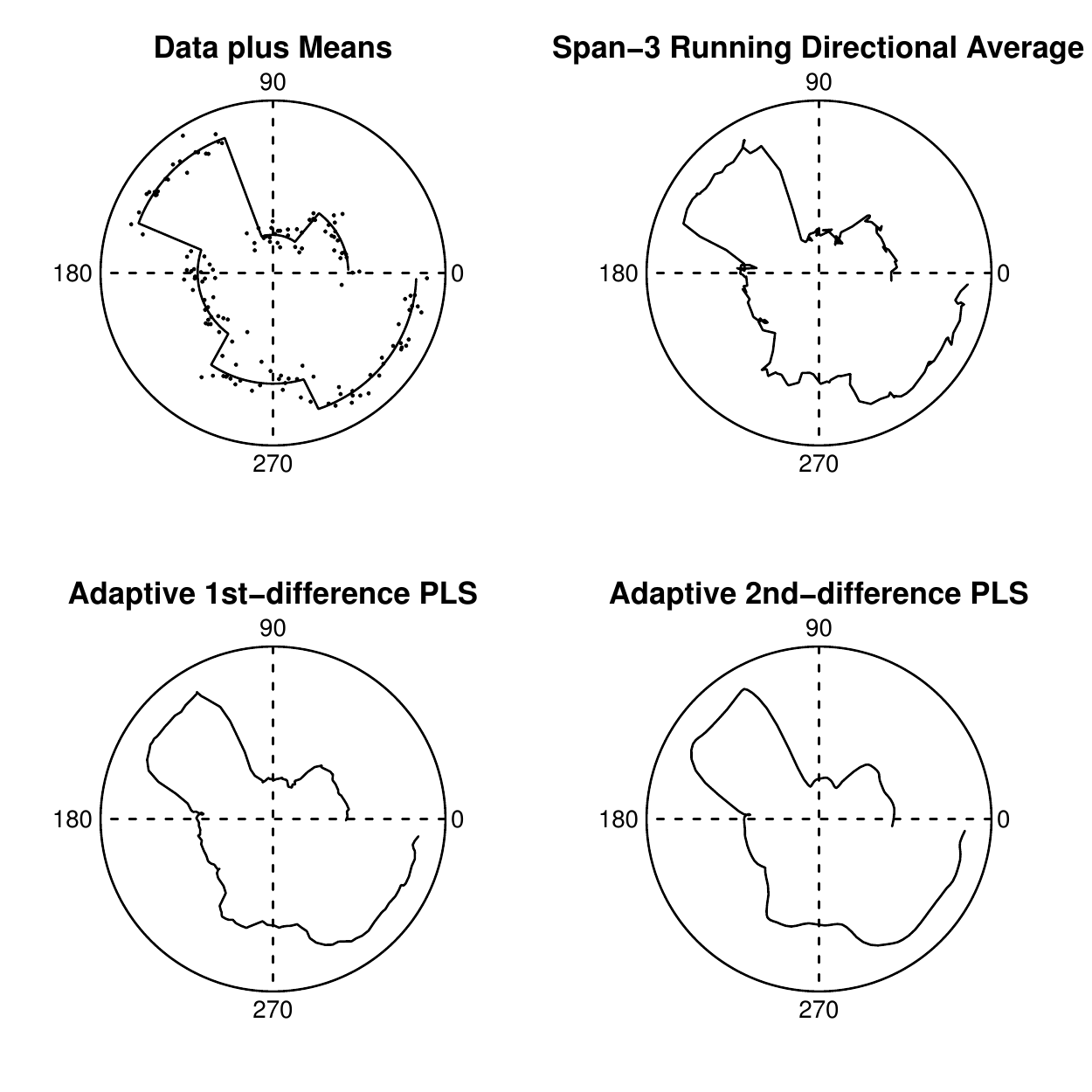}
\caption{\sl The Jumps artificial data, together with the underlying true
mean directions, plus three competing fits to the data. Interpolating lines
indicate the time sequence of the true and the estimated mean directions.}  
\label{Fig4}
\end{center}
\end{figure}

Three fits were computed for each artificial data set: the span-3 running
average, the adaptive 1st-difference PLS estimator, and the adaptive 
2nd-difference PLS estimator (cf.\ subsection \ref{4.1}). The estimator
$\gammahat^2$ defined in \eqref{3.12} was used to compute estimated risks. 

Fig.\ \ref{Fig2} displays the true means and the competing fits to the
Wobble data, with linear interpolation added solely to guide the eye. The
task is estimation of \textsl{discrete} successive mean directions observed
with error. From smallest to largest, the estimated risks of the competing
directional trend estimators are: $.0020$ (running average); $.0022$
(adaptive 2nd-difference PLS); $.0032$ (adaptive 1st-difference PLS). These
relatively similar estimated risks contrast with the much larger estimated
risk $.0138$ for the naive estimator consisting of the observed directions
$\{y_i\}$. To the eye, the smoothed fits vary in details but not in major
features.

Fig.\ \ref{Fig3} presents the true means and the competing fits to the Bat
data, again with linear interpolation added solely to guide the eye. From
smallest to largest, the estimated risks of the competing directional trend
estimators are: $.0001$ (adaptive 2nd-difference PLS); $.0018$ (adaptive
1st-difference PLS); $.0023$ (running average). These values contrast with
the much larger estimated risk $.0117$ for the naive estimator. Visually,
the adaptive 2nd-difference PLS estimator is clearly the best while the
1st-difference PLS estimator seems slightly better than the running
average.  

Fig.\ \ref{Fig4} gives the true means and the competing fits to the Jumps
data. From smallest to largest, the estimated risks of the competing
directional trend estimators are: $.0028$ (adaptive 2nd-difference PLS);
$.0031$ (adaptive 1st-difference PLS; $.0038$ (running average). These
values contrast with the much larger estimated risk $.0155$ for the naive
estimator. To the eye, the 2nd-difference PLS fit exhibits the fewest
extraneous artifacts. All three fitting techniques face difficulty in
identifying the sharp transitions.
 
\begin{figure}[ht]
\begin{center}
\includegraphics[height=2in]{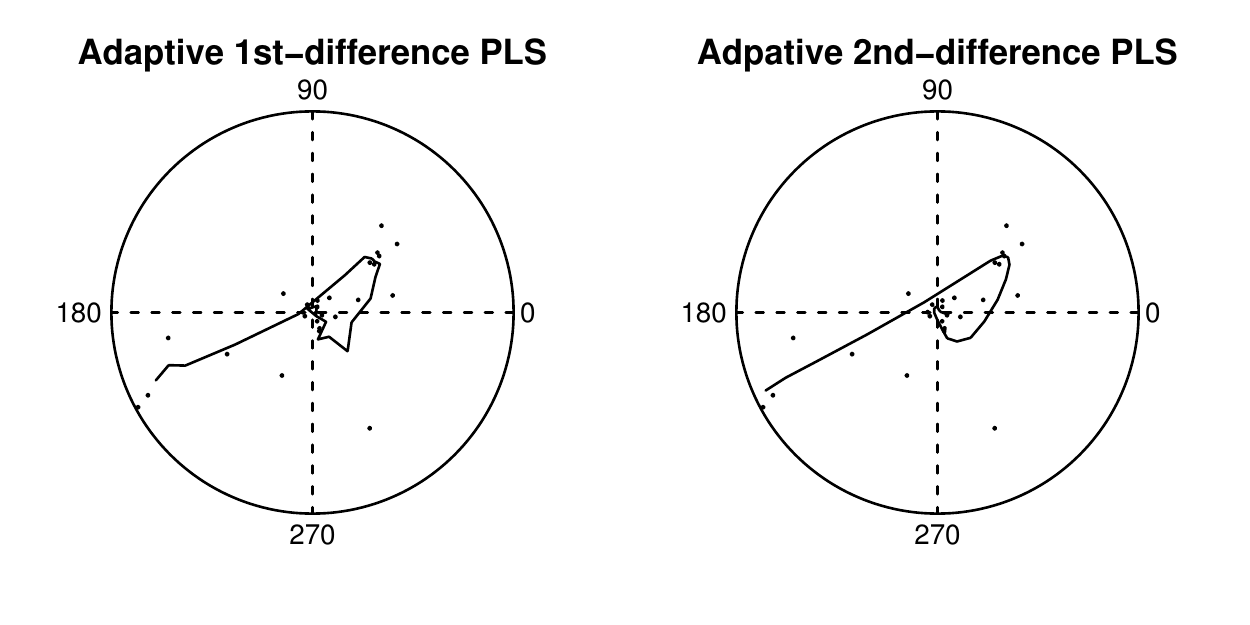}
\caption{\sl  The paleomagentic north pole data with the first- and
second-difference adaptive PLS estimated mean directions. Interpolating
lines indicate the time sequence of the fitted mean directions.}  
\label{Fig5}
\end{center}
\end{figure}

\subsection{Fitting the Paleomagnetic North Pole Data}\label{subsec4.4}
Fig.\ \ref{Fig5} exhibits the adaptive first- and second-difference PLS
fits to the sequence of paleomagnetic north pole observations described in
the Introduction. Again, linear interpolation has been added merely to
guide the eye between successive estimated mean directions. These fits are
to be compared with the raw data and with the 3-year running average
directions displayed in Fig.\ref{Fig1}. From smallest to largest, the
estimated risks of the competing directional trend estimators are: $.0158$
(adaptive 2nd-difference PLS); $.0230$ (adaptive 1st-difference PLS;
$.0303$ (running average). These values contrast with the much larger
estimated risk $.1055$ for the naive estimator consisting of the raw data. 
The adaptive 2nd-difference PLS estimator---the competitor with smallest
estimated risk---is arguably the most pleasing to the eye. It tells a clear
story of the paleomagnetic north pole spiraling, with time, to the vicinity
of the Earth's geographic north pole.

The remark of \citet{JT} on methodologies, quoted in the Introduction,
remains pertinent. The fitting procedures treated in this paper are scored
credibly by their estimated risks under the nonparametric data model of
subsection \ref{2.1}---the asymptotic theory and numerical trials in
sections \ref{sec3} and \ref{sec4} establish this point. The directional
trend fits obtained for the paleomagnetic data are highly suggestive
scientifically. Nevertheless, no certainties follow. There is no intrinsic
reason for observational errors in the paleomagnetic data to obey a
probability model. 

\section{Proofs}\label{sec5}

\subsection{Ancillary Results}\label{subsec5.1}
The theorem proofs draw on standard properties of the spectral and
Frobenius matrix norms and on two preliminary Propositions. The first
of these, from the theory of weak convergence in $C[0,1]^k$, is restated
here for convenient reference in subsequent argument. \citet{HL} summarizes
pertinent results on the relations among matrix norms. 

\begin{proposition}\label{prop1}
Let the  $\{U_p\colon p \ge 1\}$ be random elements of $C([0,1]^k)$. Let 
$\plim_{p \rightarrow \infty}$ denote the limit on probability as $p
\rightarrow\infty$. Suppose that
\begin{align}
&\plim_{p \rightarrow\infty}U_p(t) = 0 \qquad \forall t \in [0,1]^k,
\notag\\
&\lim_{\delta \rightarrow 0} \limsup_{p \rightarrow\infty} \,\P[\sup_{|t-s|
\le \delta}|U_p(t) - U_p(s)| \ge \epsilon] = 0 \qquad \forall \epsilon > 0. 
\label{5.1}
\end{align}
Then $\plim_{p \rightarrow\infty} \sup_{t \in [0,1]^k}|U_p(t)| = 0$.
\end{proposition}

The next Proposition presents the core result underlying the proof
of Theorem \ref{thm3}. The following notation and assumptions are required:

\begin{itemize}
\item
For $j=1,2$, the $p \times p$ symmetric matrices $\{C_j(t) \colon t \in
[0,1]^k\}$ are continuous on $[0,1]^k$ with $\sup_p\sup_{t \in
[0,1]^k}|C_j(t)|_{sp} < \infty$.

\item
For $j=1,2$, the matrix $C_j(t)$ is differentiable on the interior of
$[0,1]^k$, with partial derivatives  $\{\nabla_i C_j(t)
= \partial C_j(t)/\partial t_i \colon 1 \le i \le k\}$ such that
$\sup_{p,i}\sup_{t \in [0,1]^k} |\nabla_i C_j(t)|_{sp} < \infty$.

\item
The random vector $z = (z_1, z_2, \ldots, z_p)'$ has $\E(z) =
0$ and $\Cov(z) = K = \{\diag (k_{ii})\}$. The components of $z$ are
independent random variables such that $\sup_{1 \le i \le p}|z_i| \le 1$
w.p.1. 

\end{itemize}

\begin{proposition}\label{prop2}
Suppose these assumptions hold.
Let $b = (b_1, b_2, \ldots, b_p)'$ be any constant vector such that
$|b_i| \le 1$ for $1 \le i \le p$. Let
\be
U_p(t) = p^{-1}[b'C_1(t)z + \{z'C_2(t)z - \tr(C_2(t)K)\}]. 
\label{5.2}
\ee
Then
\be
\lim_{p \rightarrow \infty} \E[\sup_{t \in [0,1]^k}|U_p(t)|] = 0.
\label{5.3}
\ee
\end{proposition}

\noindent\textbf{Proof.}
\textsl{Pointwise consistency}. For notational simplicity, write $C_j$ for
$C_j(t)$ in the calculations. From the assumptions, $\E(b'C_1z)
= \E[z'C_2z - \tr(C_2)K] = 0$, $k_{ii} = \Var(z_i) \le 1$, $\Var(z_i^2) \le
1$ and $|b|^2 \le p$. Consequently,
\be
\Var(p^{-1}b'C_1 z) = p^{-2}b'C_1 K C_1 b \le p^{-2}|b|^2|C_1 K C_1|_{sp}
\le p^{-1}|C_1|_{sp}^2 .
\label{5.4}
\ee
\
Moreover, if $C_2 = \{c_{2,ij}\}$, then $z'C_2z = \sum_i c_{2,ii}z_i^2 +
2\sum_{i<j}c_{2,ij}z_i z_j$ and
\be
\Var(p^{-1}z'C_2 z) \le p^{-2}(\sum_i c_{2,ii}^2 + 4\sum_{i<j} c_{2,ij}^2)  
\le 2p^{-2}|C_2|^2 \le 2p^{-1}|C_2|_{sp}^2.
\label{5.5}
\ee
The upper bounds in \eqref{5.4} and \eqref{5.5} both tend as zero as $p$
increases. Hence,
\be
\plim_{p \rightarrow\infty}U_p(t) = 0 \qquad \forall t \in [0,1]^k.
\label{5.6}
\ee

\textsl{Uniform consistency}. For every $s,t \in [0,1]^k$,
\be
U_p(s)-U_p(t) = p^{-1}\sum_{i=1}^k (s_i-t_i)[b'\nabla_iC_1 z +
\{z'\nabla_iC_2 z - \tr(\nabla_iC_2 K)\}],
\label{5.7}
\ee
where $\nabla_iC_j = \nabla_iC_j(\bar s)$ for some $\bar s$ on the line
segment that joins $s$ and $t$. Therefore,
\be
\sup_{|s-t|\le \delta}|U_p(s)-U_p(t)| \le \delta p^{-1}\sum_{i=1}^k
[|b'\nabla_iC_1 z| + |z'\nabla_iC_2 z| + |\tr(\nabla_iC_2 K)|].
\label{5.8}
\ee
Because $|K^{1/2}|_{sp} \le 1$, $\E|z|^2 \le p$, and $|b|^2 \le p$,
\begin{align}
p^{-1}\E|b'\nabla_iC_1 z| &\le p^{-1}|b|\,\E|z|\, |\nabla_iC_1|_{sp} \le
|\nabla_iC_1|_{sp},\notag \\
p^{-1}\E|z'\nabla_iC_2 z| &\le p^{-1}\E|z|^2\, |\nabla_iC_2|_{sp} \le
|\nabla_iC_2|_{sp}, \notag \\ 
p^{-1}|\tr(K^{1/2}\,\nabla_iC_2\, K^{1/2})| &\le |K^{1/2}\,\nabla_iC_2\,
K^{1/2}|_{sp} \le |\nabla_iC_2|_{sp} .
\label{5.9}
\end{align}
In view of \eqref{5.8}, display \eqref{5.9}, and the Proposition
assumptions, Markov's inequality establishes existence of a finite constant
$c$, not depending on $p$, such that
\be
\P[\sup_{|s-t|\le \delta}|U_p(s)-U_p(t)| \ge \epsilon] \le c\delta.
\label{5.10}
\ee
Consequently,
\be
\lim_{\delta \rightarrow 0} \limsup_{p \rightarrow\infty} \,\P[\sup_{|t-s|
\le \delta}|U_p(t) - U_p(s)| \ge \epsilon] = 0. 
\label{5.11}
\ee
Limits \eqref{5.6} and \eqref{5.11}, together with Proposition \ref{prop1},
establish
\be
\plim_{p \rightarrow\infty} \sup_{t \in [0,1]^k} |U_p(t)| = 0.
\label{5.12}
\ee

\textsl{$L_1$ uniform consistency}. Observe that
\be
\sup_{t \in [0,1]^k} |U_p(t)| \le p^{-1} \sup_{t \in [0,1]^k} [|b'C_1(t) z|
+ |z'C_2(t) z| + |\tr(C_2(t) K)|]
\label{5.13}
\ee
Because $|b|^2 \le p$, $|z|^2 \le p$ w.p.1, and $|K^{1/2}|_{sp} \le 1$,
\begin{align}
p^{-1}|b'C_1(t) z| &\le |C_1(t)|_{sp}, \qquad p^{-1}|z'C_2(t) z| \le
|C_2(t)|_{sp}, \notag \\
p^{-1}|\tr(C_2(t) K)| &= p^{-1}|\tr(K^{1/2}C_2(t) K^{1/2})| \le
|C_2(t)|_{sp}.
\label{5.14}
\end{align}
Thus, the sequence $\{\sup_{t \in [0,1]^k} |U_p(t)| \colon p \ge 1\}$ is
bounded and \eqref{5.12} may be strengthened to the desired $L_1$
convergence \eqref{5.3}.\qed 

\subsection{Theorem Proofs}\label{subsec5.2}

\noindent
\textbf{Proof of Theorem \ref{thm1}.}
In view of \eqref{2.6}, \eqref{2.7}, and the symmetry of $A$, the risk is
\begin{align}
p^{-1}\E |AY-M|^2 &= p^{-1}\sum_{j=1}^q \E|Ay_{(j)} - m_{(j)}|^2 \notag \\
&= p^{-1}\sum_{j=1}^q [\tr (A^2)\Cov(y_{(j)})) + \tr((I_p - A)^2
m_{(j)}m'_{(j)}) \notag \\ 
&= p^{-1}[\gamma^2\tr(A^2) +  \tr((I_p -A)^2MM')]. 
\label{5.15}
\end{align}
This establishes \eqref{2.8}.

The spectral representation \eqref{2.9} for $A$ entails
\begin{align}
\gamma^2\tr(A^2) &= \sum_{k=1}^s \gamma^2 a_k^2\tr(P_k) \notag \\
\tr((I_p-A)^2MM') &= \sum_{k=1}^s (1-a_k)^2 |P_kM|^2.
\label{5.16}
\end{align}
Applying this to \eqref{2.8} yields the first expression in \eqref{2.10}.
Completing the square then gives the second line in \eqref{2.10}. \qed

\medskip\noindent
\textbf{Proof of Theorem \ref{thm2}}. The expression \eqref{2.12} for
$\Rhat(A)$ entails
\begin{align}
\Rhat(A) &= p^{-1}[\tr((I_p-A)^2YY') + \gammahat^2(\tr(A^2) - \tr((I_p
-A)^2)] \notag \\
&= p^{-1}[|Y-AY|^2 + \tr(2A - I_p)\gammahat^2] ,
\label{5.17}
\end{align}
which yields the desired expression \eqref{2.13}.

Arguing as in \eqref{5.16} yields
\be
\gammahat^2\tr(A^2) = \sum_{k=1}^s \gammahat^2 a_k^2\tr(P_k), \,\,\,\,\,
\tr((I_p-A)^2YY') = \sum_{k=1}^s (1-a_k)^2 |P_kY|^2
\label{5.18}
\ee
and
\be
\gammahat^2\tr((I_p-A)^2) = \sum_{k=1}^s \gammahat^2(1-a_k)^2\tr(P_k).
\label{5.19}
\ee
Substituting \eqref{5.18} and \eqref{5.19} into the first line in
\eqref{5.17} yields the first expression in \eqref{2.14}. Completing the
square for the summands with $\what_k \ge 0$ then gives the second line in
\eqref{2.14}. \qed

\medskip\noindent
\textbf{Proof of Theorem \ref{thm3}}.
The theorem assumptions stated in subsection \ref{subsec3.1} imply the
following properties for $B(t) = A^2(t)$ and $\Bbar(t) = (I_p - A(t))^2$: 
\begin{align}
\sup_p\sup_{t \in [0,1]^k} |B(t)|_{sp} < \infty, &\qquad
\sup_{p,i}\sup_{t \in [0,1]^k} |\nabla_i B(t)|_{sp} < \infty, \notag \\ 
\sup_p\sup_{t \in [0,1]^k} |\Bbar(t)|_{sp} < \infty, &\qquad
\sup_{p,i}\sup_{t \in [0,1]^k} |\nabla_i \Bbar(t)|_{sp} < \infty.
\label{5.20}
\end{align}
The bounds on derivatives use the identity $\nabla_i B(t) = \nabla_i
A(t)\cdot A(t) + A(t) \cdot \nabla_i A(t)$. 

{\sl The case $W(A) = \Rhat(A)$}. Define $\Rbreve(A)$ by replacing
$\gammahat^2$ with $\gamma^2$ in expression \eqref{2.12} for
$\Rhat(A)$. For simplicity, the argument $t$ is omitted in the
following calculations, where $B, \Bbar$ stand for $B(t), \Bbar(t)$
respectively and so forth. Because 
\begin{align}
|\Rhat(A) - \Rbreve(A)| &\le p^{-1}|\gammahat^2 - \gamma^2|(|\tr(B)| +
|\tr(\Bbar)|) \notag \\ 
&\le |\gammahat^2 - \gamma^2|(|B|_{sp} + |\Bbar|_{sp})
\label{5.21}
\end{align}
and $\gammahat^2$ is $L_1$ consistent by \eqref{3.1}, it suffices to prove
the case $W(A) = \Rbreve(A)$.

The random vector  $u_{(j)} = y_{(j)} - m_{(j)}$  has $\E(u_{(j)}) = 0$.
From \eqref{2.4}, \eqref{2.5} and \eqref{2.6}, $V_j = \Cov(u_{(j)})$ is a
diagonal matrix such that $\sum_{j=1}^q V_j = \gamma^2 I_p$. Because each
row of $Y$ and of $M$ is a unit vector, the components of $u_{(j)} =
\{u_{ij}\}$ are independent random variables such that $\sup_{1 \le i \le
p}|u_{ij}| \le 2$ w.p.1. 

Expression \eqref{2.8} for $R(\Dhat(A),M,\gamma^2)$ and the definition
above of $\Rbreve(A)$ yield
\begin{align}
\Rbreve(A) - &R(\Dhat(A),M,\gamma^2) = p^{-1}\tr[\Bbar\{\sum_{j=1}^q
(y_{(j)}y_{(j)}'- m_{(j)}m_{(j)}') - \gamma^2 I_p\}] \notag \\
&= p^{-1}\tr[\Bbar\sum_{j=1}^q (u_{(j)}u_{(j)}' + m_{(j)}u_{(j)}' +
u_{(j)}m_{(j)}' - V_j)] \notag \\
&= p^{-1}\sum_{j=1}^q[2m_{(j)}'\Bbar u_{(j)} + \{u_{(j)}'\Bbar u_{(j)} -
\tr(\Bbar V_j)\}].
\label{5.22}
\end{align}
Apply Proposition \ref{prop2} separately to each summand on the
right side of \eqref{5.22} after setting $z = u_{(j)}/2$, $b = m_{(j)}$,
$C_1(t) = C_2(t)= 4\Bbar(t)$, and $K = V_j/4$. This completes the proof of
\eqref{3.7} when $W(A) = \Rhat(A)$. 

{\sl The case $W(A) = L(\Dhat(A),M)$}. The argument is similar. Let $F =
A^2 - A$, a symmetric matrix. Note that the loss defined in \eqref{2.7} may
be rewritten
\begin{align}
L(\Dhat(A),M) &= p^{-1}\sum_{j=1}^q |Au_{(j)} + (A - I_p)m_{(j)}|^2 \notag
\\
&= p^{-1}\sum_{j=1}^q [2m_{(j)}'Fu_{(j)} + u_{(j)}'B u_{(j)} +
m_{(j)}'\Bbar m_{(j)}].
\label{5.23}
\end{align}
On the other hand, the risk \eqref{2.8} may be rewritten
\be
R(\Dhat(A),M,\gamma^2) = p^{-1}\sum_{j=1}^q [\tr(BV_j) + 
m_{(j)}'\Bbar m_{(j)}].
\label{5.24}
\ee
Consequently, $\Delta = L(\Dhat(A),M) - R(\Dhat(A),M,\gamma^2)$ satisfies 
\be
\Delta = p^{-1}\sum_{j=1}^q
[2m_{(j)}'Fu_{(j)} + \{u_{(j)}'B u_{(j)} - \tr(BV_j)\}].
\label{5.25}
\ee
Apply Proposition \ref{prop2} separately to each summand on the
right side of \eqref{5.25} after setting $z = u_{(j)}/2$, $b = m_{(j)}$,
$C_1(t) = 4(A(t)^2 - A(t))$, $C_2(t)= 4B(t)$, and $K = V_j/4$. This
completes the proof of \eqref{3.7} when $W(A) = L(\Dhat(A),M)$. \qed 

\medskip\noindent
\textbf{Proof of Theorem \ref{thm4}}.
Let $r(A,M,\gamma^2) = R(\Dhat(A),M,\gamma^2)$. It suffices to show that
\eqref{3.7} in Theorem \ref{thm3} implies
\be
\lim_{p \rightarrow\infty} \sup_{M} \E |Z -r(\Atilde,M,\gamma^2)| =
0,
\label{5.26}
\ee
where $Z$ is either the loss of $\Dhat(\Ahat)$ or the estimated risk
$\Rhat(\Ahat)$. The three limits to be proved in \eqref{3.9} and
\eqref{3.10} are then immediate.

First, \eqref{3.7} with $W(A) = \Rhat(A)$ implies
\begin{align}
\lim_{p \rightarrow\infty} \sup_{M} \E |\Rhat(\Ahat) - r(\Ahat, M,
\gamma^2)| &= 0, \notag \\
\lim_{p \rightarrow\infty} \sup_{M} \E |\Rhat(\Ahat) - r(\Atilde,
M, \gamma^2)| &= 0.
\label{5.27}
\end{align}
Consequently, \eqref{5.26} holds for $Z = \Rhat(\Ahat)$ and
\be
\lim_{p \rightarrow\infty} \sup_{M} \E | r(\Ahat, M, \gamma^2)-
r(\Atilde,M, \gamma^2)| = 0.
\label{5.28}
\ee

Second, \eqref{3.7} with $W(A) = L(\Dhat(A),M)$ yields
\be
\lim_{p \rightarrow\infty} \sup_{M} \E |L(\Dhat(\Ahat),M) -
r(\Ahat,M,\gamma^2)| = 0.
\label{5.29}
\ee
This and \eqref{5.28} establish that \eqref{5.26} holds for $Z =
L(\Dhat(\Ahat),M)$. \qed

\medskip\noindent
\textbf{Proof of Theorem \ref{thm5}}. By algebraic expansion of
\eqref{3.12},
\be
\gammahat^2 = [2(p-1)]^{-1}\sum_{i=2}^p [|y_i-m_i|^2 +
|y_{i-1}-m_{i-1}|^2] + |m_i-m_{i-1}|^2] +T_p,
\label{5.30}
\ee
where $(p-1)T_p$ is the quantity
\be
\sum_{i=2}^p [(y_i-m_i)'(m_i-m_{i-1}) - (y_{i-1}-m_{i-1})'
(m_i-m_{i-1}) - (y_i-m_i)'(y_{i-1}-m_{i-1})'].
\label{5.31}
\ee
Recall from Section \ref{sec2} that $|y_i| = |m_i| = 1$, $\E (y_i) = m_i$,
and $\E |y_i - m_i|^2 = \tr(\Sigma) = \gamma^2$ for each $i$.
Straightforward variance calculations plus Chebyshev's inequality establish
that 
\begin{align}
\plim_{p \rightarrow\infty} (p-1)^{-1}\sum_{i=2}^p |y_i-m_i|^2 &=
\gamma^2 = 
\plim_{p \rightarrow\infty} (p-1)^{-1}\sum_{i=2}^p |y_{i-1}-m_{i-1}|^2
\notag \\ 
\plim_{p \rightarrow\infty} T_p &= 0. 
\end{align}
This together with the assumption \eqref{3.11} establishes $\plim_{p
\rightarrow\infty} \gammahat^2 = \gamma^2$. This may be strengthened to
$L_1$ convergence because each of the normalized sums above is bounded. A
closer look at the argument verifies the desired uniformity in $M$. \qed

\end{document}